\documentclass[twocolumn,showpacs,preprintnumbers,amsmath,amssymb,pre]{revtex4}

\usepackage{graphics}
\usepackage{amsmath}
\usepackage{graphicx,color}
\usepackage[latin1]{inputenc}
\usepackage[T1]{fontenc}
\usepackage{ae,aecompl}

\begin{document}

\title{Correlation length for amorphous systems}
\author{Jorge Kurchan}

\affiliation{PMMH-CNRS UMR 7636-ESPCI-Universit\'e Paris 6-Universit\'e Paris 7, rue Vauquelin 10, 75231 Paris, France}
\author{Dov Levine }

\affiliation{Department of Physics, Technion, Haifa 32000, Israel}


\begin{abstract}
{Crystals and quasicrystals  can be  characterized by an
 order that is a  purely geometric property of  an  instantaneous configuration, independent of  particle dynamics or interactions. 
Glasses, on the other hand, are ostensibly amorphous arrangements of particles. A  natural and 
 long-standing question has been whether they too have, albeit in a hidden way,   some  form of geometric order.
 Here we define a coherence  length that  
  applies to systems which  are typically characterized as amorphous, as well  as to those that are conventionally ordered.  We argue that the divergence of such a length is consistent with
current theories of the  `ideal' glass transition. 
}
\end{abstract}
\maketitle

There comes a moment in many  introductory talks on glasses when 
 two pictures are projected showing configurations of a liquid and  a glass, 
and the speaker
defies the audience to tell which is which.  
The purpose of this paper is to propose a 
purely geometric correlation length that may
indeed do precisely this,  characterizing  glass structure  even though we may not know, a priori, what
the ideal glass state is. The idea is to consider the spatial frequency with which any given patch of linear size  $\ell$ and volume ${V} \sim \ell^d$ appears in an infinite sample. 
 We shall argue in what follows  that  patches smaller than a certain  size $\ell_o$ occur `often' --  with an average repeat distance ${\cal D}$ that scales  sub-exponentially with their volume, 
 while
 patches larger than $\ell_o$ do so `rarely', with ${\cal D}$ being exponential in $V$, {\it i.e.,}  just as one would expect of a random system.
The existence  of  an `ideal glass state' then involves a divergence of this crossover length  $\ell_o$ (with a concomitant diverging time scale) below the thermodynamic transition temperature $T_K$,
implying that in such a  state motifs of {\em all} sizes recur frequently.
We note that our definition makes no reference to either 
interactions or equilibration, nor is it restricted to systems of a specific type of constituent ({\it e.g.,} particle, polymer, or continuous density).
 In the rest of the paper we shall proceed, in the context of several examples, in two steps:
first, we argue for the existence of an ordered $\ell_o=\infty$ phase where all patterns recur frequently, and second, we describe a mechanism whereby this length becomes finite.


That some form of amorphous order exists in the glass phase is implicit in several theories of fragile glasses, most notably 
in the `Random First Order' scenario~\cite{RFOT, Cavagna}. One assumes that  there is some `ideal glass' state 
to which the system would equilibrate, given an infinitely slow annealing from high temperature to 
 $T < T_{K}$, where $T_{K}$ is the Kauzmann temperature.  Such an ideal state should be characterised by permanent
non-periodic spatial density-modulations.
 In this ideal glass, the entropy, or, more precisely, the {\em complexity} (the logarithm of the number of average density profiles)~\cite{footnote} is subextensive in the size of the system.
The main point of this paper is that the further assumption that in the ideal glass state the complexity   of patches of volume ${V}$ which may be found in an infinite system~\cite{foot0}
 is  subextensive  may be used to define a 
purely geometric length scale. To illustrate this point, we shall discuss, in  some detail, a more stringent variant of this constraint, in which the state of a
macroscopic portion of a sample determines, if not completely, at
least significantly, the adjacent region. 

 A thought experiment illustrating
this~\cite{Franz_Parisi,Kob_Parisi,Biroli_Bouchaud,Cavagna_Grigera}  proceeds as follows: First, equilibrate  a large sample and choose a region $\Sigma$ of linear size $\ell$ and volume $V\propto \ell^{d}$.  Denote the (microscopic) configuration of this region by $A$, and look at another region $\Sigma'$, contiguous with $\Sigma$, whose volume $V' = \alpha V$ is proportional to that of  $V$;  denote its configuration by $A'$.  Next, melt the entire sample except for $\Sigma$, which remains frozen in configuration $A$.  Now, allow the rest of the system to equilibrate; the region $\Sigma'$ will now solidify into a new configuration $B$.  If (with some probability) $B \sim A'$ in some metrical sense, this means
that the configuration in $\Sigma$ determines (to some extent) that of a neighboring region $\Sigma'$ proportional to its own size. Such a scenario is
 expected for quenches to $T<T_K$~\cite{MS}. 

\begin{figure}[htbp]
\begin{center}
\vspace{-.8cm}
\includegraphics[width=3cm,angle=+0]{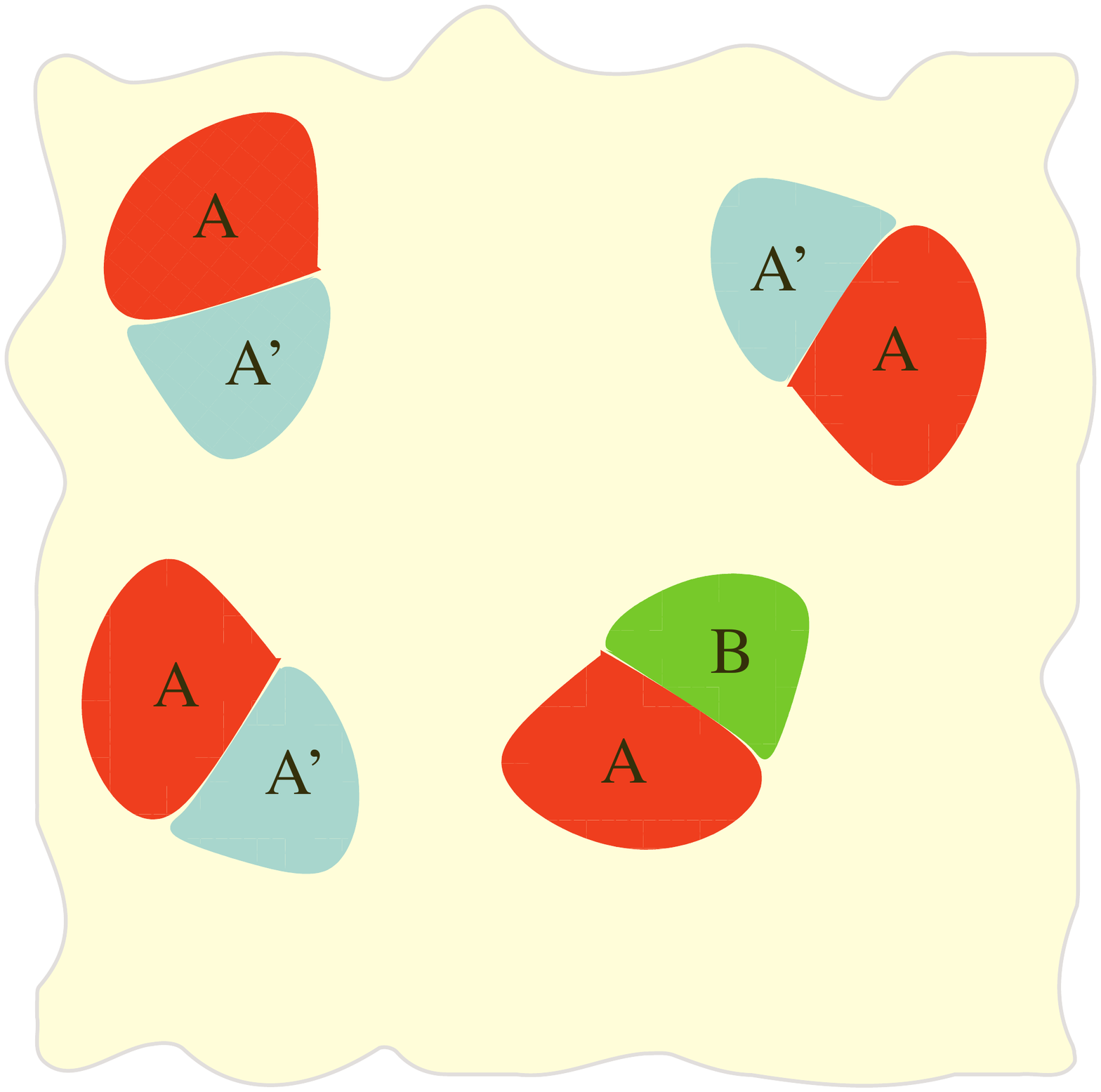}
\end{center}
\vspace{-1cm}
\caption{Sub-extensivity implies static correlation: The surroundings of identical patches are constrained. }
\label{fig1}
\end{figure}

To motivate the definition of our correlation length, consider the following static variant of the preceding thought experiment.  Beginning with an infinitely large configuration of a perfect equilibrium ({\it i.e.,} ideal)
glass, mark all regions of size $\ell$ isomorphic to $\Sigma$, that is, whose configuration is $A$; 
this is indicated shematically in Fig. \ref{fig1} (we consider only statistically isotropic systems here).  We will loosely refer to all these regions as $\Sigma$'s.
The configurations of regions $\Sigma'$ of volume $\alpha V$ contiguous with one of the $\Sigma$'s  
is (with finite  probability) determined by that of the neighboring $\Sigma$.  Noting that $\Sigma$ is representative of regions of size $\ell$,
it follows that
if $\Omega(V)$ is the number of possible configurations that a region of volume $V$ may
take, then the constraint that $\Sigma$ determines a finite fraction of $\Sigma\Sigma'$ pairs 
may be expressed as: 
\begin{equation}
\Omega(\alpha {V}) = \gamma\; \Omega({V})\;\;\;\;.
\end{equation}
If $\gamma$ is either constant or varies slowly with $V$, we have that
\begin{equation}
\Omega(V) \sim V^{\mu d}
\label{powerlaw}
\end{equation}
with $\mu = \frac{\ln \gamma}{d \ln \alpha}$.  We note that for the case of a periodic arrangement,
the state of $\Sigma'$ is completely determined, so $\gamma = 1$, implying $\mu=0$. 
On the other extreme, if the ideal glass state is random, then patches are  independent and we have that $\ln \Omega(  {V}) \propto V$: 
the complexity is proportional to the volume.
 Assuming 
that equal size patches occur with comparable  probability, the distance ${\cal D}(\ell)$ we must 
go before we happen upon an identical patch of size ${V}\sim\ell^d$ goes as ${\cal D}(\ell)\sim \Omega^{1/d}(V)$.  
In the  example (\ref{powerlaw}) above, ${\cal{D}}$ scales as power law  ${\cal{D}} \sim \ell^\mu$, to be contrasted with the uncorrelated case where
 it scales exponentially. 
We emphasize that the assumption that a region of volume ${V}$ significantly restricts the possibilities of an 
adjacent region of size $\alpha {V}$ (with the concomitant power law for ${\cal{D}}$)  is rather strong. Although we regard this scenario as physically compelling,
our definition of a diverging correlation length is more general, as it  only requires that the number  of embedded density profiles is subexponential in  ${V}$, since then  the repetition distance ${\cal{D}}$  will  scale with $\ell$ 
subexponentially, which is all we need.

With the previous discussion of pattern repetition in the ideal glass, we are now poised to define the coherence length.  Given an (infinitely large) snapshot of 
a  sample, we pick a pattern of size
$\ell$ and measure the spatial distance ${\cal D}(\ell)$ with which it
appears to within a specified precision,
and repeat this procedure for the different patches of size $\ell$ (coincidence may be defined modulo rotations, but this is inessential).  Performing
this procedure for different sizes, we plot, say, the average
${\cal D}(\ell)$ as a function of $\ell$. If the sample is above $T_K$, or if it is below $T_K$  but has not fully equilibrated to the ideal glass state, only small-scale 
patterns will repeat
frequently, with large patterns recurring infrequently.  That is, at some length $\ell_{o}$ there will be a crossover from subexponential behavior ({\it e.g.;} ${\cal D} \sim \ell^\mu$),
to a regime where the repetitions are much more
sparse:  ${\cal D} \sim e^{a \ell^d}$. We define 
$\ell_o$  as the coherence length.

We may relax the rigid requirement  of strict identity and decide that two regions are 
congruent even if there is a number of mismatches. This   allows us to work at  
finite temperatures, and to discuss correlation lengths in systems undergoing
general phase transitions.  For example, in the case of  a ferromagnet, the correct prescription is to  identify two patterns if their overlap is larger or equal than 
the magnetization squared.    One may also consider the generalization to coarse-grained, or
even topological descriptions, which have recently proved interesting in the study of
glassy systems\cite{Procaccia}.




It is important to note that sub-extensive complexity of embedded patches,  such as that of Equation (\ref{powerlaw}), does not imply that the infinite system is either periodic or quasiperiodic (these are characterized by a diffraction spectrum consisting of  $\delta-$function Bragg peaks). To make this clear, we
discuss a particularly well understood  class of systems, namely those which possess `recursive symmetry'.  Examples of such systems are one-dimensional sequences of two symbols (1 and 0) which may be generated by means
of a substitution rule, wherein a symbol is replaced by a specific word (composed
of 0's and 1's) everywhere it occurs; generalization to higher 
dimensions is immediate.  We consider three prototypes
which characterize three types of spatial structure:
\begin{equation}
{\bf Rule \;1:} \; \; \; \; \; \; \; \; \; \; \; \; 1 \rightarrow 10  \; \; \; \; \; \; 0 \rightarrow 10
\end{equation}
Starting from (say) 0, we generate sequences of length $2^{n}$ by iterating 
this rule $n$ times; the result is clearly
$$
0101010101010101010101010101
$$
Continuing in this fashion, we generate a simple periodic sequence of arbitrary
length.  The second possibility is illustrated by 
\begin{equation}
{\bf Rule \;2:} \; \; \; \; \; \; \; \; \; \; \; \; 1 \rightarrow 10  \; \; \; \; \; \; 0 \rightarrow 1
\end{equation}
If we start from 0 as before, we generate strings of length $F_{n}$, where
$F_{n}$ is the $n^{th}$ Fibonacci number:
$$
1011010110110101101011011010110110
$$
This rule produces the {\it Fibonacci sequence}, which is quasiperiodic, 
whose Fourier transfom consists of a dense set of $\delta$-functions.
Finally, we consider 
\begin{equation}
{\bf Rule\; 3:} \; \; \; \; \; \; \; \; \; \; \; \; 1 \rightarrow 10  \; \; \; \; \; \; 0 \rightarrow 01
\end{equation}
Iterating this rule $n$ times produces a string of length $2^{n}$:
$$
01101001100101101001011001101001
$$
This sequence, known as the {\it Thue-Morse sequence}, has no  $\delta$-functions
in its Fourier transform~\cite{Bombieri,thue-morse,Godreche_Luck}. 
This kind of sequence also differs from  quasiperiodic ones in that there are no {\em finite} translations that
take a  {\em macroscopic} system into a similar configuration~\cite{Aubry}. 


Consider the multiplicity of substrings of length $\ell$ of an infinite sequence  generated by the above rules:
in the first case there are only two possibilities, and in the last two 
$O(\ell)$~\cite{fib,berthe}:
all  patches repeat in a distance ${\cal{D}}$  proportional to their size $\ell$. To see this, note that a subsequence  of size $\ell$ is completely included within the `descendants' of a single site after $k \propto \log(\ell)$ inflation steps; the  subsequence repeats in the descendant of a neighbouring ancestor -
its `cousin $k$-times removed'. {\em This argument is applicable to all systems obtained by substitutions.}
  We conclude that  the coherence length $\ell_o$ is infinite for a perfect sequence -- even though there may be no Bragg peaks in the diffraction spectrum, so that by the conventional measure of scattering such systems would be considered amorphous.

\begin{figure}[htbp]
\includegraphics[width=5cm]{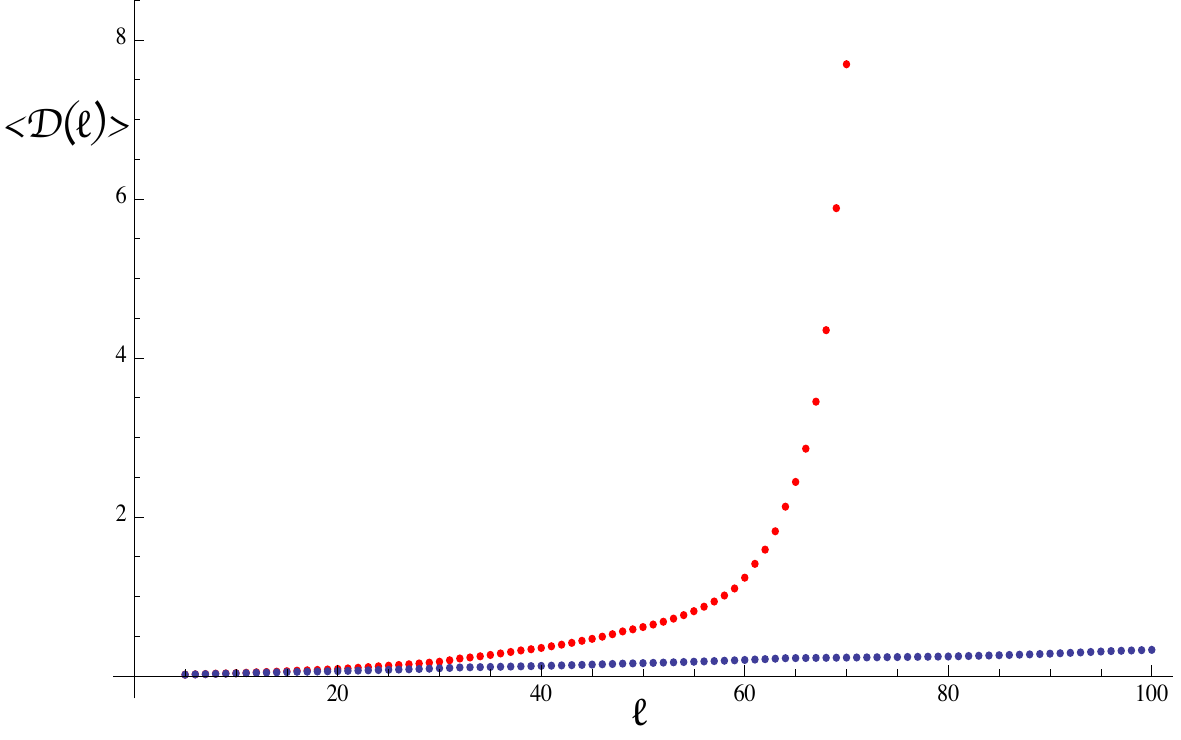}
\caption{ The average distance ${\cal D}(\ell)$ between words of length $\ell$ in 
a perfect Thue-Morse sequence of length 1000 (blue) and a sequence of length 1000 composed by concatenating
strings of average length 50 randomly drawn from a perfect Thue-Morse sequence
(red).}
\label{avedist}
\end{figure}
%
%
%
%

Next, in analogy to a system which has not achieved the ideal glass state, we consider a sequence constructed by concatenating unrelated subsequences of varying lengths $n$, drawn from a distribution $P(n)$, with average $\overline{n}$.  It is easy to see that strings of length $\ell \ll \overline{n}$ will repeat often, because even in independent subsequences the same string will appear.  In this case,
${\cal D}(\ell)$ will remain close to its value in the ideal sequence.
However, if $\ell \gg \overline{n}$, the likelihood of overlap is exponentially small, by the central limit theorem.  This is seen in Figure \ref{avedist},
where the average distance between strings of length $\ell$ is plotted as a function of $\ell$ for a perfect and a randomly concatenated Thue-Morse sequence.

\begin{figure}[htbp]
\begin{center}
\includegraphics[width=6cm]{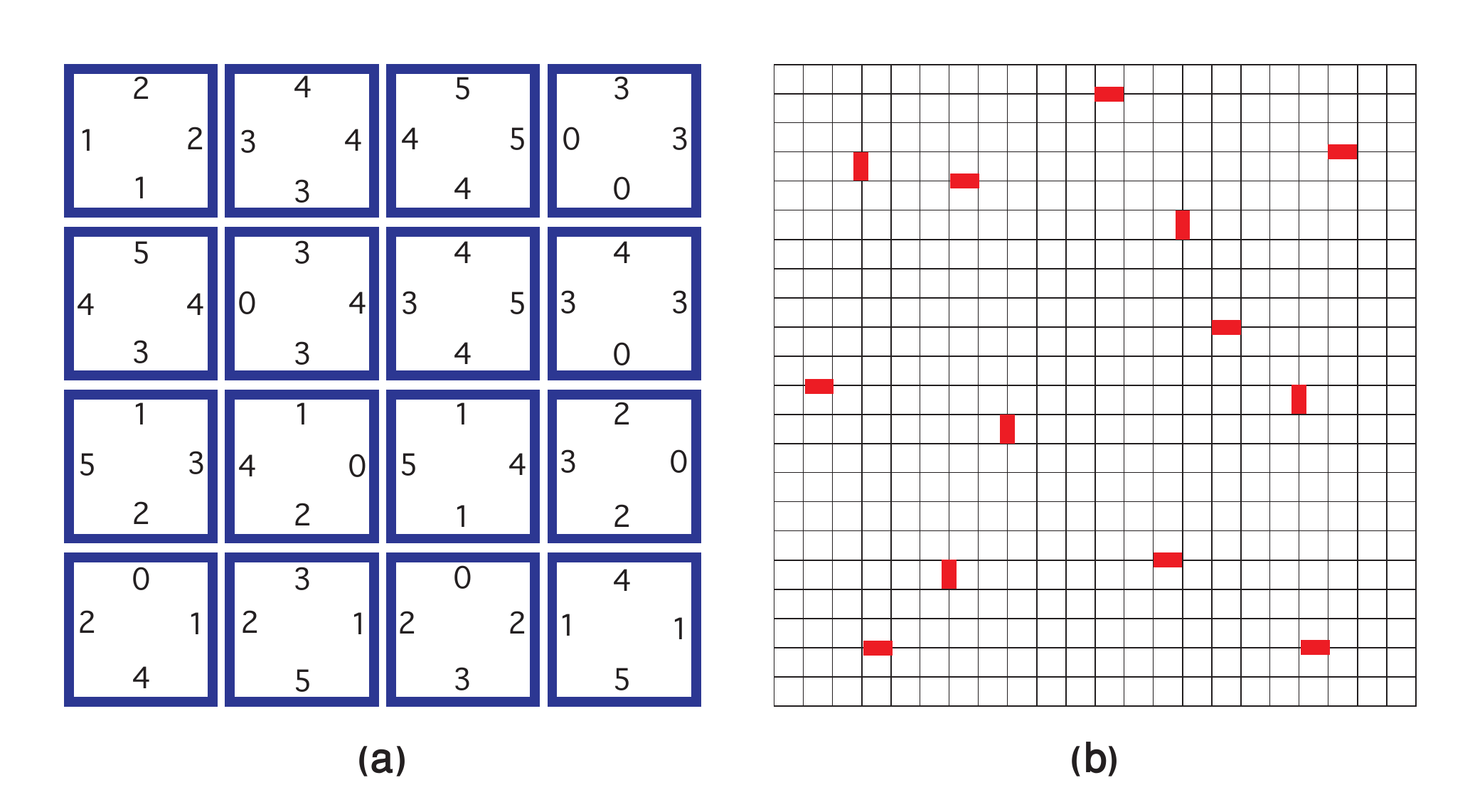}
\end{center}
\vspace{-.5cm}
\caption{(a) A set of Wang tiles with matching rules indicated by numbered edges.  The rule for laying the tiles is that only edges of the same number may abut. (b) A state from a Monte-Carlo simulation of a Wang tiling quenched from high to low temperature: Only the
`hard' defects  ({\it i.e.} mismatches of adjacent sides) remain; they are indicated by red lines.}
\label{wangtiles}
\end{figure}

These considerations may be generalized for higher dimensions as well.
Perhaps the best known 2D quasiperiodic tiling is the Penrose tiling, but a system with similar properties is a Wang tiling~\cite{Grunbaum}.  In constructing a Wang tiling by successively laying tiles one next to another as a mason would,
one is required to follow matching rules.  One set of Wang tiles with their associated matching rules is shown in Figure \ref{wangtiles}(a). 
An example of a system  that is statistically isotropic is a tiling of identical triangles~\cite{Radin} 
called the 'Pinwheel' tiling.
The Pinwheel tiling can be constructed by recursive substitution rules, so patterns of
size $\ell$ repeat (modulo rotation) every ${\cal D} \sim \ell$ in a perfect tiling, as for the 1D sequences 
discussed above.  Significantly, it is known that any packing which can be generated 
by a recursive substitution rule admits a set of matching rules which only allows
tilings consistent with the recursive symmetry~\cite{Goodman-Srauss}; the
set of matching rules for the Pinwheel tiling is known explicitly~\cite{Radin}.

%
\begin{figure}[htbp]
\includegraphics[width=4cm]{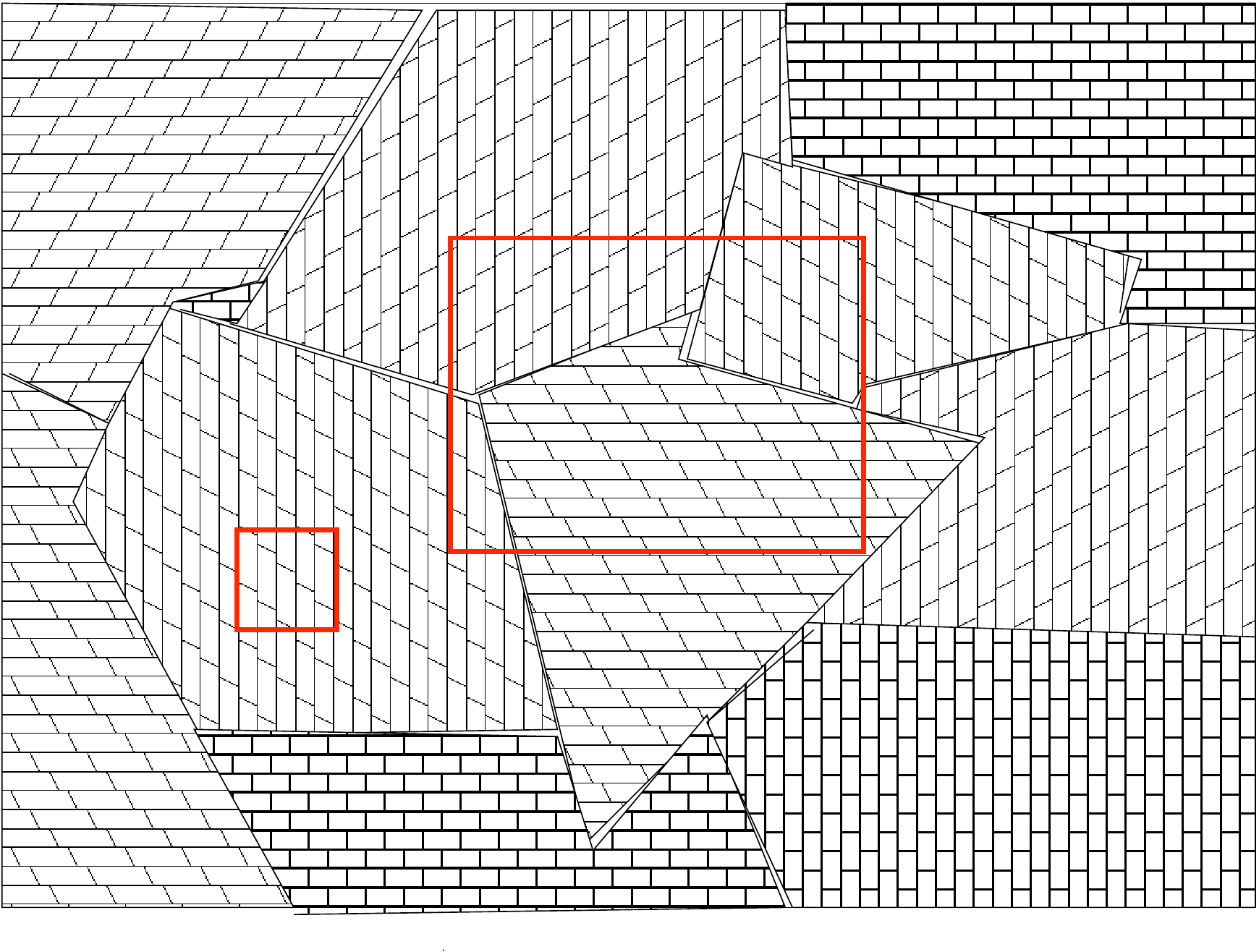}
\caption{ Two patterns, one smaller and one larger than  the crystallite length. 
  The small pattern will recur with high  frequency, but the large
  one will be exponentially rare, because it requires recurrence of the crystallite boundaries within a patch, which is very unlikely.}
\label{mosaic1}
\vspace{-.5cm}
\end{figure}

Having discussed the prefect order case $\ell_o=\infty$, we need to  understand the nature of the defects that would reduce
$\ell_{o}$.
Figure \ref{mosaic1}
illustrates how this comes about in the  polycrystalline case;
clearly $\ell_o$ measures the typical crystallite size.  
While in this simple case the decoherence which decreases $\ell_{o}$ is effected by well-defined grain boundaries, for general systems decoherence may occur due to less obvious defects. 
Consider for example the  Wang tiling, which  has the virtue that it can 
be expressed as a lattice model of spins~\cite{Leuzzi} (taking $16$ values), assigning an energy corresponding to a mismatched edge - that is to
say, the matching rules may be enforced by a suitable Hamiltonian. 
Consequently, such a system can be simulated  at finite temperature (See Figure \ref{wangtiles}(b)) to measure  the growth of the coherence length.
At finite temperature, the Wang tile system has energetic {\it point} 
defects - local violations of the matching rules.
Starting from a high temperature state and annealing down to low temperatures, many easy-to-cure `benign' defects disappear,
leaving the  system in a state with isolated `hard' defects (Figure \ref{wangtiles}(b)) that 
can only be removed by large-scale rearrangements of the order of the interdefect
distance.  Such behavior is also present in the Penrose tilings, where there are benign 
matching rule violations due to local tile flips, and non-trivial configurations of the 
phason field~\cite{coherence} which require large-scale rearrangements to eradicate.  In this case, as we
have verified numerically for the Wang tile system, the
correlation length $\ell_o$ is of the order of the distance between `hard' defects.   
The conclusion we draw from this is that  grain boundaries, playing both the role
 of borders of coherent regions  
and simultaneously being the place where energy is locally
concentrated, is a peculiarity of crystals. Further work is necessary to elucidate the various mechanisms for loss of coherence in quasicrystals, and nothing is known about such mechanisms for `amorphous' recursive systems.

It is worth relating the interesting concept of `hyperunifomity' (or `superhomogeneity'), applied in  the context of amorphous
sphere packings~\cite{Torquato,Joyce}, with the current work.  A system
is said to be hyperuniform if the square fluctuations in particle number between regions of identical volume $V$ scales more slowly than $V$.  
One can restate this  in terms of recurrence of   
patches of volume $V$ having the same particle number,  
while the  correlation length  defined here entails  recurrence of  actual  {\it patterns}~\cite{Joyce}. 
Note that  a random stacking of hexagonal layers of hard
spheres
is hyperuniform, but at least in the direction normal to the layers, it has a finite $\ell_{o}$.




 


 The proposal that $\ell_o$ diverges in the glass phase may seem to contradict the  notion that glasses are 
 `spatially chaotic',  since truly chaotic solutions do not have the frequent pattern repetitions we are claiming here.
We argue that repetitions only appear  when one considers the {\em lowest} free-energy states. A similar  
 situation was already encountered in charge density waves ~\cite{Aubry,CF}, where chaos reduces to quasiperiodicity as soon as one restricts to the lowest energy solutions.


To conclude, one may make the strong and intriguing conjecture that ideal glassy states have some recursive symmetry that can be constructed with some (yet unknown) inflation rule applying  to continuous densities.  In this case, it would be the emergence of this symmetry which is responsible for the diverging time scale.


We wish to thank L. Berthier, G. Biroli, S. Franz, Y. Kafri, N. Merhav, M. Sellitto and F. Zamponi for illuminating discussions. DL 
gratefully acknowledges support from the Israel Science Foundation, under Grant 1574/08.



\begin{thebibliography}{100}
\bibitem{footnote}  More precisely, the complexity can be  defined
as the logarithm of the number of configurations, but identifying   all configurations leading in finite times to the 
same time-averaged density profile. It can also be implemented in practice by identifying configurations that differ less than a conveniently chosen threshold value. 
 For example, the entropy of a simple crystal includes contributions from phonons, etc, whereas the complexity is zero, 
since there is but a single crystal state.
\bibitem{foot0} This is not the usual definition of  configurational entropy, where  one takes the volume $V$  with fixed boundary conditions.
Here we are considering that $V$ is  embedded in an infinite sample, so that it is influenced by the state of its surroundings. This may
increase the number of possible density profiles,  by a factor
at most equal to the multiplicity of  possible states of  the  boundary, a quantity  sub-exponential in $V$.  
Conversely, a  ground state for an isolated volume $V$ may be impossible to continue
outside $V$ without defects, and thus be  discarded when $V$ is embedded in a large thermalised state.
\end{thebibliography}
\end{document}